\documentclass[showpacs,twocolumn,preprintnumbers,amsmath,amssymb,pra]{revtex4}

\usepackage{graphicx}
\usepackage{dcolumn}
\usepackage{bm}
\usepackage{psfrag}
\usepackage{epic}
\usepackage{eepic}
\usepackage{bm}
\usepackage{mathrsfs}

\newcommand{\trace}[1]{\mathrm{Tr}\, #1}

\newcommand{\bra}[1]{{\langle #1|}}
\newcommand{\ket}[1]{{|#1 \rangle}}
\newcommand{\braket}[2]{{\langle #1|#2 \rangle}}
\newcommand{\expt}[1]{\left\langle #1\right\rangle}

\newcommand{\imi}{\ensuremath{\mathrm{i}}}

\begin{document}

\newenvironment{smatrix}{\begin{bmatrix}}{\end{bmatrix}}
\setlength{\tabcolsep}{4ex}
\setlength{\unitlength}{1mm}

\title{Entanglement properties of quantum spin chains}

\author{Stein Olav Skr{\o}vseth}
\email{stein.skrovseth@phys.ntnu.no}
\affiliation{%
Department of Physics,
Norwegian University of Science and Technology,
N-7491 Trondheim, Norway
}%

\date{1 March 2006}

\begin{abstract}
  We investigate the entanglement properties of a finite size 1+1
  dimensional Ising spin
  chain, and show how these properties scale and can be utilized to
  reconstruct the ground state wave function. Even at the critical
  point, few terms in a Schmidt decomposition contribute to the exact
  ground state, and to physical properties such as the
  entropy. Nevertheless the entanglement here is prominent due to the
  lower-lying states in the Schmidt decomposition.
\end{abstract}


\maketitle

\section{Introduction}
The EPR argument \cite{EPR35} and the Bell inequalities
\cite{Bell64} were, albeit with almost 30  
years in between, acknowledgments
that quantum theory exhibits the strange correlations known as
entanglement. However, only in the latest few years has it been
known that entanglement is a resource that can be utilized in quantum
computing,\cite{Nielsen&Chuang} and is thus a central subject in the
continually expanding field of 
quantum information theory. Furthermore, entanglement has been shown
to be a fundamental feature in quantum phase transitions, something
that has spawned a whole new field of research \cite{Sachdev,
  Osterloh:2002, Vidal:2002rm, SOS1, Osborne:2002, Korepin2004, SOSbose,
  Calabrese04, calabrese-2005-0504, latorre05}. This article will 
focus on entanglement in both quantum critical and non critical
systems, the structure of entanglement in condensed matter systems
being at its most complex in critical systems. 

The determination of the ground state in a quantum system such as 1D
spin chains is highly complex, but the last years have seen the
development of techniques such as DMRG \cite{White1992,Schollwoeck2005}
and the recent entanglement renormalization \cite{Vidal2005} which
solve the problem efficiently. In both cases the entanglement in the
problem seems to be the key to the methods' success over traditional
renormalization schemes. Moreover, for a pure state, any long range
correlations require the existence of entanglement in the state.

A quantum system with wave function $\ket\Psi$ in a Hilbert space
$\mathcal H$, which is partitioned
into two subspaces $\mathcal H_{\mathscr A}\otimes\mathcal H_{\mathscr
B}$, can be written as a Schmidt decomposition 
\begin{equation}
  \ket\Psi=\sum_{n=1}^\chi\sqrt{\lambda_n}\,
  \ket{\psi_n^{(\mathscr A)}}\otimes\ket{\psi_n^{(\mathscr B)}},
  \label{Schmidt}
\end{equation}
where $\ket{\psi_n^{(\mathscr A)}}\in\mathcal H_{\mathscr A}$ and vice
versa. The coefficients $\lambda_n$ are real, positive c-numbers and
the states are mutually orthogonal, $\braket{\psi_n^{\mathscr
    A}}{\psi_m^{\mathscr A}}=\delta_{nm}$. The upper
limit $\chi$, the Schmidt number, is a brute measure of the
entanglement between the two subsystems, running from one (no
entanglement), maximally reaching the dimensionality of the smallest
of the two Hilbert spaces. The reduced density matrix of (say) system
$\mathscr A$ can then be written
\[\rho^{(\mathscr A)}=\trace_{\mathscr
  B}\ket\Psi\bra\Psi=\sum_{n=0}^\chi\lambda_n\ket{\psi_n^{(\mathscr
    A)}}\bra{\psi_n^{(\mathscr A)}}\]
and equivalently for $\mathscr B$. Hence, knowing the Schmidt
decomposition of the wave function is equivalent to knowing the basis
in which the reduced density matrix is diagonal, modulo phases. We
will in this 
article show that the effective Schmidt number $\chi_{\mathrm{eff}}$
is much smaller than the possible maximum, and thus the number of
terms contributing significantly to the wave function is
surprisingly small. Remarkably, this also is true at highly entangled
points in the phase space, such as at criticality. 

The entropy of the wave function $\ket\Psi$ can be measured as the entropy
of the subsystem,
\begin{equation}
  S_{\mathscr A}=-\trace \rho^{(\mathscr A)}\log_2\rho^{(\mathscr A)}=-\sum_n\lambda_n\log_2\lambda_n,
  \label{Sdef}
\end{equation}
increasing from zero if the two subspaces are entangled. Hence,
computing the
eigenvalues $\lambda_n$ of the reduced density matrix is vital to the
understanding of entangled states.

\section{Ising model}
The quantum Ising chain in external field $h$ in 1+1 dimension is a
good benchmark tool for 
the analysis of entanglement since its properties are extremely
well known. The model is defined by the Hamiltonian 
\begin{equation}
  H_{\mathrm{Ising}}=-\sum_{n=1}^N\left(\sigma_n^x\sigma_{n+1}^x+h\sigma_n^z\right),
\end{equation}
on $N$ lattice sites.
We operate with open boundary conditions (OBC), since this will
ease the formalities when partitioning the system. This will weaken
the phase transition in the sense that the conformal symmetry is
broken compared to periodic boundary conditions (PBC)
\cite{Francesco97}, but the phase 
transition in the thermodynamic limit will prevail. Also, Calabrese
and Cardy have found that one can compute conformal signatures in the
OBC case provided the boundary conditions are conformal \cite{Calabrese04}.
The model has a phase transition at $h=1$ between the product state
$\ket{\Psi_\infty}=\ket{\uparrow\uparrow\cdots\uparrow\,}$ for $h>1$ and
the Schr\"{o}dinger cat state
$\ket{\Psi_0}=\frac1{\sqrt2}\left(\ket{\rightarrow\rightarrow\cdots\rightarrow}+\ket{\leftarrow\leftarrow\cdots\leftarrow}\right)$ 
when $h<1$. Here $\ket{\!\!\uparrow}$ is the eigenstate of $\sigma^z$
with eigenvalue 1, and $\ket{\!\!\leftrightarrows}$ are
eigenstates of $\sigma^x$ with eigenvalues $\pm1$. Note that the
transition is between unit entropy in the low field limit and zero in
the high field limit. 

Doing a Jordan-Wigner transform as sketched in Refs. \cite{Latorre:2003kg,
  Osborne:2002}, we can map the model onto a string of non interacting
fermions, and thus compute the eigenvalues for the Ising model (and
a larger class of models) in what resembles the thermodynamic
limit, typically a few hundred particles. In essence, we define the
$N$ fermionic operators 
\begin{equation}
  \hat
  a_n=\frac12\left(\bigotimes_{k=1}^{n-1}\sigma_k^z\right)\otimes\left(\sigma_n^x+\imi\sigma_n^y\right),
  \label{def_fermions}
\end{equation}
and the $2N$ Majorana fermions 
\[\check\gamma_{2n}=\frac1{\imi\sqrt2}\left(\hat a_n-\hat
  a_n^\dag\right),\quad
\check\gamma_{2n-1}=\frac1{\sqrt2}\left(\hat a_n+\hat a_n^\dag\right).\]
These fulfill $\{\check\gamma_m,\check\gamma_n\}=\delta_{mn}$, and
are delocalized in terms of the original lattice of fermions. The
Majorana fermions
diagonalize the Hamiltonian in the sense that
\[H=\sum_{mn}C_{mn}\check\gamma_m\check\gamma_n,\]
with an Hermitian $2N\times2N$ matrix $C$. Next, define the imaginary
and anti symmetric correlation matrix 
$\Gamma_{ij}=\expt{[\check\gamma_i,\check\gamma_j]}$. We will consider
only the ground state, and thus the expectation
values are to be taken in the ground state. In this sense, our model is
defined by the matrix $C$ while the state is defined in the matrix
$\Gamma$. The Majorana fermions have a two-to-one correspondence to
the fermions (\ref{def_fermions}), and hence tracing out a particle
from the system amounts to removing the two adjacent rows and columns
in $\Gamma$ corresponding to the particle. Tracing out $N-N'$
particles this way, we end up with the $2N'\times2N'$ matrix
$\bar\Gamma_{ij}$. This 
latter matrix can be made block diagonal through an orthogonal
transformation $O$, such that we can define
new Majorana fermions $\bar\gamma_n$ that fulfill
\begin{equation}
  O^{\mathrm T}\bar\Gamma O=\left\{\expt{[\bar\gamma_i,\bar\gamma_j]}\right\}=\bigoplus_{k=1}^{N'}\begin{pmatrix}0&\imi\xi_k\\-\imi\xi_k&0\end{pmatrix}
  \label{bargammadef}
\end{equation}
with $0\leq \xi_k\leq1$. The transformation $O$ is the same
transformation that block diagonalizes the Hamiltonian matrix $C$.
The matrix $\bar\Gamma$ corresponds to the
state described by the reduced density matrix $\rho'$, with the $N-N'$
particles traced out. The fermion 
operators $\bar a_n$ corresponding to $\bar\gamma_n$ will
diagonalize the reduced density matrix such that the eigenvalues
thereof can be computed. The eigenvalues are
determined by the set of 
binary occupation numbers $\eta=\{n_k\}$, $k=1,2,\ldots,N'$, and
$n_k=\{0,1\}$, $\ket\eta$ being an eigenstate of $\bar a_k\bar
a_k^\dag$ with 
eigenvalue $n_k$. Thus $\ket\eta$ is also an eigenstate of $\rho'$ and
the reduced density matrix is
$\rho'=\sum_\eta\lambda_\eta\ket\eta\bra\eta$ with eigenvalues 
\begin{equation}
  \lambda_\eta=\prod_{k=1}^{N'}\left[\frac12+(-1)^{n_k}\xi_k\right].
  \label{lambdafromxi}
\end{equation}
Finally, the entropy becomes 
\begin{equation}
  S_{\mathscr A}=\sum_{k=1}^N H\left(\frac12\left(1+\xi_k\right)\right),
\end{equation}
where $H(x)=-x\log_2x-(1-x)\log_2(1-x)$ is the binary entropy function. 

\begin{figure}[htbp]
  \includegraphics[width=.9\columnwidth]{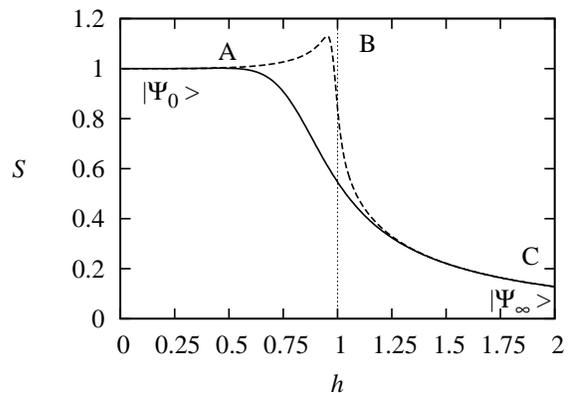}
  \caption{The entropy in the quantum Ising model as function of the
    external magnetic field $h$ for $N=10$ (lower line) and $N=100$
    particles. In both cases half system is traced out, from the
    edge. The letters refer to the places of investigation in 
    Fig. \ref{fig:eigvalues_Ising}. The critical point (in the
    thermodynamic limit) is shown as a vertical line.} 
  \label{fig:SIsing}
\end{figure}

The entropy of the ground state as measured by
Eq. (\ref{Sdef}) when the system is partitioned into two equal parts,
is shown in Fig. \ref{fig:SIsing}.
The increased entropy around the critical point is a hallmark of the
quantum phase transition, though this does not show in the small system.

\begin{figure}[htbp]
  \includegraphics[width=.9\columnwidth]{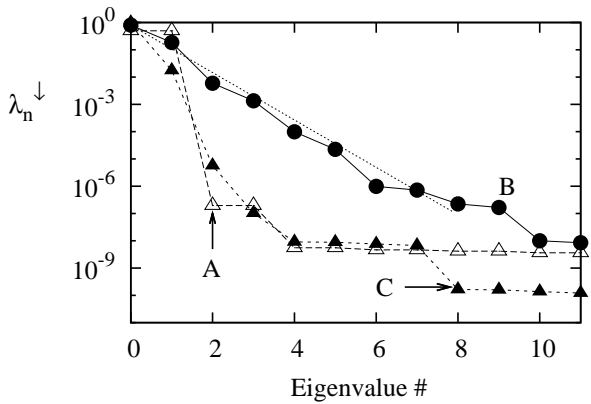}
  \caption{The decreasingly ordered eigenvalues $\lambda_n^\downarrow$
    of the reduced density 
    matrix of the Ising model's ground state when tracing out half
    size. Data for $N=50$. The upper line ($\bullet$) denotes the critical values
    when $h=h_c=1$, at (B) in figure \ref{fig:SIsing}; ($\triangle$)
    denotes the Scr\"{o}dinger cat state with $h=0.5$, at (A); and
    ($\blacktriangle$) denotes the approximate product state with $h=2$, at
    (C). The fitting line to the critical values
    $\lambda_n^\downarrow\sim e^{-2n}$ is shown. 
    Only the 10 largest eigenvalues are shown since numerical
    errors become prominent after this.}
  \label{fig:eigvalues_Ising}
\end{figure}
Figure \ref{fig:eigvalues_Ising} shows the magnitude of the
decreasingly ordered 
eigenvalues $\lambda_n^\downarrow$. Note that there are $2^{N/2}=128$ possible
eigenvalue contributions, while this shows that only a very few of
these contribute to the wave function, and thus to the entanglement as
measured by the entropy $S$.

The main eigenvalues at critical point decrease roughly exponentially,
as $\lambda_n^\downarrow\sim e^{-2n}$, while off criticality only the
very first contribute significantly.
In the low field limit, the 
highest eigenvalues are paired due to the Schr\"{o}dinger cat nature
of the state, while in the high field limit there is one single main
contribution, which is the product state $\ket{\Psi_\infty}$. With
PBC, the eigenvalues would be paired also for $h>1$, except for the
single main eigenvalue due to the translational symmetry. This pairing
also occurs in a conformal bosonic chain \cite{SOSbose}, but is not a
general property of a non critical system.

Hence we can approximate the wave function by restricting the sum
(\ref{Schmidt}) to some 
upper limit $\chi'<\chi$ to make the new wave function 
\[\ket{\Psi'}=\sum_{n=1}^{\chi'}\sqrt{\lambda_n'}\ket{\psi_n^{(\mathscr
    A)}}\otimes\ket{\psi_n^{(\mathscr B)}},\]
where the new coefficients $\lambda_n'$ are determined by
normalization, $\lambda_n'=\lambda_n/(1-\varepsilon)$. 
We define an error as
\[\varepsilon=\sum_{n=\chi'+1}^\chi\lambda_n\]
which measures the difference in the eigenvalue sum of the two
wave functions. The overlap becomes
\[\braket{\Psi'}\Psi=\sqrt{1-\varepsilon}.\]

\section{Scaling of eigenvalues}
We investigate how the eigenvalues scale with increasing system
size. Off the critical point there
is {\it no scaling} with system size of the eigenvalues since the
entropy saturates at some value. However, on, or near the critical
point the entanglement entropy 
diverges as predicted by conformal field theory. Given that we trace
out a constant fraction of the 
entire system, the entanglement entropy diverges up to an additive
constant as \cite{Vidal:2002rm, Calabrese04}
\[S_{\mathscr A}\sim\frac c6\log_2(N),\]
where $c$ is the central charge of the conformal field theory
corresponding to the phase transition. In the Ising case $c=1/2$.

\begin{figure}[htbp]
  \includegraphics[width=.9\columnwidth]{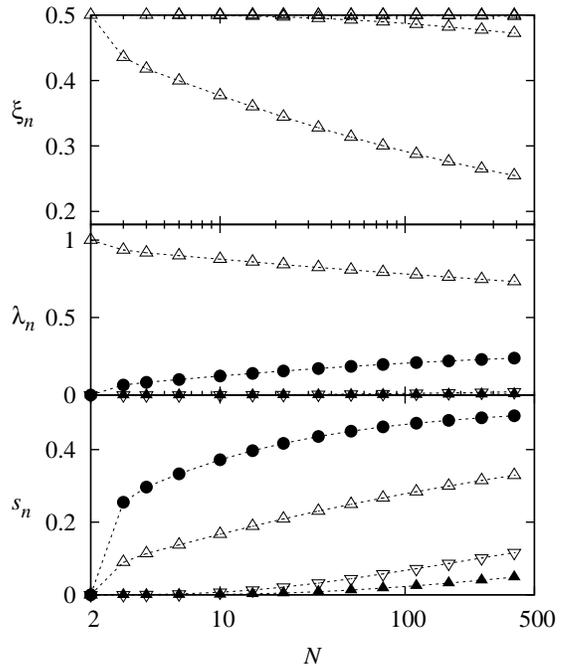}
  \caption{From the top, the figures show the four smallest
    eigenvalues of the correlation matrix in Eq. (\ref{bargammadef}),
    the 
    four largest eigenvalues of the density matrix as computed from
    Eq. (\ref{lambdafromxi}), and the corresponding entropy
    contributions  
    $s_n=-\lambda_n\log\lambda_n$. For the two lower graphs, equal
    point types refer to the same eigenvalues. System is at the
    critical Ising point and is traced at half size.}
  \label{fig:scaling_Ising}
\end{figure}
Figure
\ref{fig:scaling_Ising} shows how the largest eigenvalues scale at the
critical Ising point.
The largest eigenvalue $\lambda^\downarrow_1(N)$ {\it decreases} with system size
at criticality, while the other increases. The largest decrease
roughly as $\lambda^\downarrow_1(N)\sim-0.027\log N$, while the second increases
roughly as $\lambda^\downarrow_2(N)\sim0.021\log N$, both within the range shown
in Fig. \ref{fig:scaling_Ising}.
However, all entropy contributions increase with $N$, thus
contributing to the overall divergence of the entanglement entropy.

\section{Approximating the ground state}
There exists successful techniques to approximate the ground state of a
quantum system, the most prominent of which is the DMRG
scheme \cite{White1992}. Rather than taking only ground
states of a 
sub lattice  when renormalizing the system, this method takes into
account possible entangled states as well, producing unprecedented
accuracy. The role of entanglement in DMRG is still under
investigation, but it seems clear that it is vital to the success of
the scheme \cite{Schollwoeck2005}.
The results of our work here are indicative that the main
contributions to the ground state are indeed entangled at the critical
point, along with few of these terms contributing to the actual
wave function. These results are not directly applicable to an
improvement of the DMRG algorithm, but rather indications as to
the success of DMRG. Furthermore, any prospective technique to find
the ground state of a quantum system, even in those areas where known
techniques fail, needs to understand the nature of the wave function in
the system, and we believe that the entanglement properties would play
an important part in such a method. In particular, the entanglement
properties is precisely what distinguishes a quantum many-body system
from the classical counterpart, and entanglement
must therefore be an essential part of any such method.

\begin{figure}[htp]
  \includegraphics[width=.9\columnwidth]{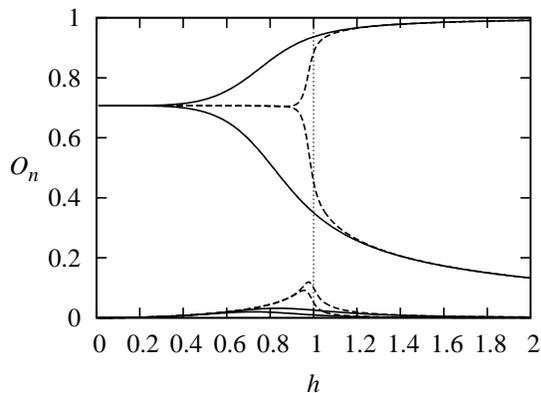}
  \caption{The four largest overlaps $\mathcal O_n$, $n=1,2,3,4$ (from
    top to bottom) with
    magnetic field $h$ and system sizes $N=10$ (full lines) and
    $N=100$ (dashed lines). Again, the critical point is indicated
    with the vertial line.}
  \label{fig:overlaps}
\end{figure}

Having found that only few terms contribute to the entanglement in the
ground state, we query how well we could possibly approximate the
ground state of the full system by the first few terms in the Schmidt
decomposition. To this end, we focus on the few-particle case, where
the exact wave functions can be computed explicitly. We split the open
spin chain in two partitions, and compute the overlap between the
actual wave function and the Schmidt expansion of the two subsystems,
ordered decreasingly on those terms with the largest overlaps.
\begin{align*}
  \mathcal O_n&=\braket{\Psi}{\psi_n^{(\mathscr
      A)}}\otimes\ket{\psi_n^{(\mathscr B)}}.
\end{align*}
{\it A priori} it is clear that in the zero field limit $\mathcal
O_1=\mathcal O_2=1/\sqrt2$, while any higher terms vanish. In the
high-field limit, $\mathcal O_1=1$ as the only non vanishing
overlap. We assume that the wave functions $\ket{\psi_n^{(\cdot)}}$
are ordered with decreasing Schmidt number, and the results are shown
in Fig. \ref{fig:overlaps}.

Since the fermions $\bar a_k$ corresponding to the Majorana fermions
defined in Eq. (\ref{bargammadef}) 
diagonalize the reduced density matrix, the overlap follows
straightforwardly. The probability of the state $\ket\eta$ given the
density matrix $\rho'=\lambda_\eta\ket\eta\bra\eta$ is $\lambda_\eta$,
and the overlap between this and the ground state will therefore be
$\sqrt{\lambda_\eta}$, or
\[\mathcal O_\eta=\prod_k\sqrt{\frac12+(-1)^{n_k}\xi_k}.\] 

Most
prominently, the $\mathcal O_1$ overlap increases monotonically as $h$
increases. That is, even at criticality this first term consisting of
the ground states at each side 
approximates the actual wave function even better that it does for the
cat-state where the overlap naturally is $1/\sqrt2$. The increased
entropy at critical points is mainly due to 
other terms rising around criticality, terms that are not part of the
entire wave function at non critical points in the phase
space.

\begin{figure}[t]
  \includegraphics[width=.9\columnwidth]{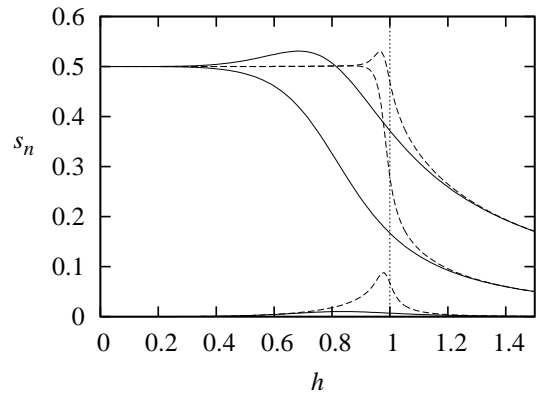}
  \caption{The three main entropy contributions $s_n$, $n=1,2,3$ (from
    top to bottom) in
    the Ising model 
    with magnetic field $h$. Full lines are for $N=10$, while dashed
    lines are for $N=100$.}
  \label{fig:coeffs}
\end{figure}

Investigating the Schmidt coefficients' entropy
contributions, the results are shown in Fig. \ref{fig:coeffs}. Hence
we see that apart from the two major contributions that are
equally prominent at $h\to0$, there is one more contributing at
criticality. The effect is mainly pronounced with big systems, but a
trace is also seen at $N=10$. The increased entropy at the critical
point is contributed to by the largest term, but also by the
appearance of terms that are zero at all noncritical points.

\begin{figure}
  \includegraphics[width=\columnwidth]{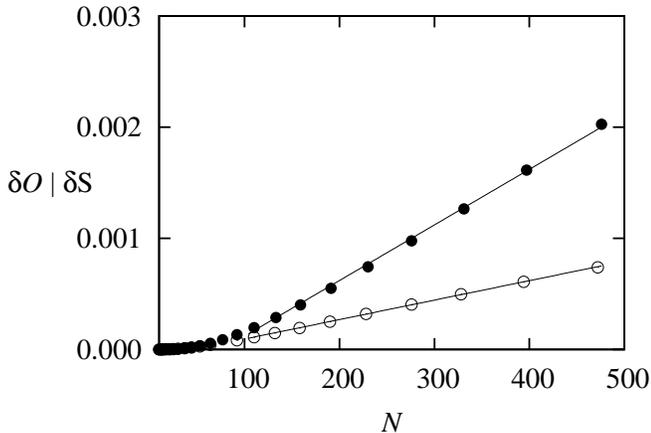}
  \caption{The errors $\delta\mathcal O$ ($\circ$) and $\delta S$
    ($\bullet$) as defined in 
    the text with different system sizes at the critical Ising
    point. For $\delta\mathcal O$ we have chosen $\chi'=4$ and for
    $\delta S$ $\chi'=3$. In all
    cases the system is traced at half-size. The lines 
    for $N>100$ are linears
    $\delta\mathcal O\simeq5\times10^{-6}N-3.8\times10^{-4}$ and $\delta
    S\simeq1.75\times10^{-6}N-8.1\times10^{-5}$. For small systems the linearity
    does not hold.}
  \label{fig:errors}
\end{figure}

The error in the overlap and in the entropy that arises from only
choosing $\chi'<\chi$ terms in the Schmidt decomposition is zero (or
very close) off criticality, since the Schmidt rank is very
small. However, on criticality $\chi$ in principle extends to the
dimension of the Hilbert space. However, the errors in the overlap
$\delta\mathcal O=1-\sum_{n=1}^{\chi'}\mathcal O_n$ and entropy $\delta
S=S-\sum_{n=1}^{\chi'}s_n$ are plotted in Fig \ref{fig:errors}. We see
that the errors grow linearly  with $N$ when $N>100$. 
The number of Schmidt terms needed to compute the entropy up to an
error of $10^{-4}$, or the effective Schmidt number is only 4 up to a
system of $N=400$ sites.

\section*{Conclusions}
We have seen that the complex structure of a wave function at critical
point in the Ising model comes from the emergence of several terms in
the Schmidt decomposition of the wave function. Nevertheless,
remarkably few terms actually contribute to the expansion, and
physical properties such as the entropy can be extracted using very
few terms, even at criticality. Mainly, the rapid decrease of the
Schmidt coefficients on or off criticality identifies this effect. We
have detailed the effects with both large and small systems, and the
primary conclusions hold in both cases.

The matter that these few eigenstates contribute to physical
properties such as the entropy and also the overlap with the true
ground state signifies the fact that finding the true ground state, or
an approximation to it, is not a complicated task {\it per se}, we
known that only a few of the eigenstates of the Hamiltonian would
suffice for a sufficiently good description of the state. The results in 
this paper also indicate that these states are indeed entangled, and
thus any algorithm that does not include entanglement is bound to
fail. 

The effective Schmidt number that is needed to find a state within
some limit is small, on the order of 10 even up to several hundred
particles. Thus the effective Hilbert space is dramatically
reduced. However, we have not addressed the problem of determining the
unknown ground state of a system, only pointed out that the problem is
not consistently hard given the correct (so far unknown) approach.

\section*{Acknowledgments}
The author is greatly indebted to K{\aa}re Olaussen for valuable
contributions to this work. Susanne Viefers and the NordForsk network
on low-dimensional physics is thanked for hosting valuable meetings.
The University of Troms{\o} (UiT{\o}) is thanked for providing
office space.

\bibliography{art}

\end{document}